\def\year{2024}\relax
\documentclass[letterpaper]{article} % DO NOT CHANGE THIS
\usepackage{aaai22}  % DO NOT CHANGE THIS
\usepackage{times}  % DO NOT CHANGE THIS
\usepackage{helvet}  % DO NOT CHANGE THIS
\usepackage{courier}  % DO NOT CHANGE THIS
\usepackage[hyphens]{url}  % DO NOT CHANGE THIS
\usepackage{graphicx} % DO NOT CHANGE THIS
\usepackage{natbib}  % DO NOT CHANGE THIS AND DO NOT ADD ANY OPTIONS TO IT
\usepackage{caption} % DO NOT CHANGE THIS AND DO NOT ADD ANY OPTIONS TO IT
\DeclareCaptionStyle{ruled}{labelfont=normalfont,labelsep=colon,strut=off} % DO NOT CHANGE THIS
\usepackage{algorithm}
\usepackage{algorithmic}
\usepackage{booktabs}
\usepackage{multirow}
\usepackage{xcolor}

\newcommand{\answerYes}[1]{\textcolor{blue}{#1}} 
\newcommand{\answerNo}[1]{\textcolor{teal}{#1}} 
\newcommand{\answerNA}[1]{\textcolor{gray}{#1}} 
 
\newcommand{\spara}[1]{\smallskip\noindent\textbf{#1.}}
\newcommand{\rev}[1]{#1}
\newcommand{\newrev}[1]{#1}

\usepackage{newfloat}
\usepackage{listings}
\floatstyle{ruled}
\newfloat{listing}{tb}{lst}{}
\floatname{listing}{Listing}

\title{Integrated or Segregated?\\ User Behavior Change after Cross-Party Interactions on Reddit}
\author{
    Yan Xia,\textsuperscript{\rm 1}
    Corrado Monti,\textsuperscript{\rm 2}
    Barbara Keller,\textsuperscript{\rm 1}
    Mikko Kivel\"{a}\textsuperscript{\rm 1}
}
\affiliations{
    \textsuperscript{\rm 1}Department of Computer Science, Aalto University, Espoo, Finland\\
    \textsuperscript{\rm 2}CENTAI, Turin, Italy\\
    yan.xia@aalto.fi, 
    corrado.monti@centai.eu, 
    barbara.keller@aalto.fi,
    mikko.kivela@aalto.fi
}

\begin{document}

\maketitle

\begin{abstract}
It has been a widely shared concern that social media reinforces echo chambers of like-minded users and exacerbate political polarization. While fostering interactions across party lines is recognized as an important strategy to break echo chambers, there is a lack of empirical evidence on whether users will actually become more integrated or instead more segregated following such interactions on real social media platforms. We fill this gap by inspecting how users change their community engagement after receiving a cross-party reply in the U.S. politics discussion on Reddit. More specifically, we investigate if they increase their activity in communities of the opposing party, or in communities of their own party. We find that receiving a cross-party reply to a comment in a non-partisan discussion space is not significantly associated with increased out-party subreddit activity, unless the comment itself is already a reply to another comment. Meanwhile, receiving a cross-party reply is significantly associated with increased in-party subreddit activity, but the effect is comparable to that of receiving a same-party reply. Our results \newrev{reveal a highly conditional depolarization effect following cross-party interactions in spurring activity in out-party communities, which is likely part of a more general dynamic of feedback-boosted engagement}.
\end{abstract}

\section{Introduction}
Political polarization has been growing in the past few decades \rev{in the U.S.} \cite{pew2017partisan,iyengar2019origins,finkel2020political}, both ideologically -- characterized by increased opinion divide on ideological issues \cite{mccarty2016polarized} -- and affectively -- characterized by increased animosity toward out-party members \cite{iyengar2012affect}. The situation is alarming as political polarization can contribute to a harsh political climate~\cite{jackson2018fake}, dehumanization~\cite{harel2020normalization}, and even outbursts of violence~\cite{michael2017rise}.

The rise of social media is seen as one of the factors that contribute to this worrying situation. In particular, researchers argue that on social media, algorithms create \emph{filter bubbles} that isolate users from content they disagree with \cite{pariser2011filter}, while users themselves form \emph{echo chambers} where they only receive information from and interact with like-minded others \cite{sunstein2018republic}.

To counter the formation and reinforcement of echo chambers on social media, researchers have proposed algorithms for fostering interactions between users of opposite political leanings \cite{garimella2017reducing, musco2018minimizing, garimella2017factors, tommasel2021want}.
Real-world applications of similar strategies were reportedly taken into consideration by Twitter in 2018.\footnote{\url{https://www.washingtonpost.com/technology/2018/08/15/jack-dorsey-says-hes-rethinking-core-how-twitter-works/}}

However, it is not clear whether these strategies will actually break echo chambers and bring polarized users together. While the intergroup contact theory \cite{allport1954nature,pettigrew1998intergroup} implies that cross-party interactions may decrease polarization (i.e., lead to depolarization) and the results of many social experiments \cite{levendusky2021we,rossiter2020consequences,combs2023reducing,de2024cross} align with this theory, some other studies \cite{bail2018exposure,wojcieszak2011deliberation} hint at the possibility that they may spark a \emph{backfire effect} -- that is, they may instead increase polarization and contribute to further isolation of opposing sides. Apart from mixed evidence from experimental studies, there is also a lack of empirical studies on how real social media users behave after interacting with members of the opposing party.

\rev{What will fill this gap is} an unobtrusive analysis of user behavior on a real social media site where cross-party interactions naturally happen. Reddit\footnote{\url{https://www.reddit.com/}}, one of the most popular news aggregation and discussion websites in the U.S., serves as an ideal platform of study, as researchers have found that the level of echo chamber effect is lower on Reddit than on Twitter and Facebook \cite{cinelli2021echo}, and heterophilic interactions between members of opposing parties are actually more common than homophilic interactions in Reddit discussions of U.S. politics \cite{defrancisci2021noecho,monti2023evidence}.

We therefore set out to study the user behavior change following cross-party interactions in Reddit discussions of U.S. politics. Aligning with the echo chamber narrative, our research revolves around the following question: after a user receives a reply from someone of the opposite political leaning, are they more likely to break out of their echo chamber and increase their activity in communities of the opposing party (i.e., \rev{\textit{out-party activity}})? Or alternatively, are they more likely to increase their activity in communities of their own party (i.e., \rev{\textit{in-party activity}})?

The temporal nature of Reddit data allows us to study the after-effect of cross-party interactions using a before-observation-after time frame. For an \textit{observation period} of \rev{two months}, we first collect all comments posted in \texttt{r/news}, the non-partisan and most popular subreddit (i.e., subcommunity) \rev{that discusses U.S. news}. \rev{Within the observation period, we then identify for each user the time they received the first reply in \texttt{r/news}, or, if they did not receive any, the time they posted the first comment in \texttt{r/news}. We consider the 30-day period before this time as the \textit{before period} and the 30-day period after as the \textit{after period}. After collecting} the comment posting history of all involved users in the before period and the after period, we infer the party affiliation of each user based on their activity in a set of partisan subreddits we identified.

We then use a regression model to inspect the association between receiving a cross-party reply in \texttt{r/news} and participating more in out-/in-party subreddits in the after period than in the before period. We analyze the robustness of our results by running the model separately on data from each of five different years and \rev{observing how consistent the results are}.

Our analysis shows that after a user receives a cross-party reply to a comment they posted in \texttt{r/news}, they are not necessarily more likely to increase their activity in out-party subreddits, unless the original comment is already a reply to another comment. Meanwhile, receiving a cross-party reply in \texttt{r/news} is significantly associated with subsequently increased in-party activity, but receiving a same-party reply is associated with a comparable effect.

Complementing previous experimental studies that measured the attitudinal or opinion change induced by cross-cutting conversation \cite{levendusky2021we,rossiter2020consequences,combs2023reducing,de2024cross}, we take a behavioral analysis perspective and reveal that \rev{in the U.S. politics discussion on Reddit,} the depolarization effect of a \rev{single} cross-party interaction in spurring the receiver's out-party community engagement is limited, unless the receiver already demonstrates a level of willingness to engage with other opinions\newrev{; and in the latter case, the increase in out-party community engagement can be driven by a more general motivating effect of receiving peer feedback \cite{joyce2006predicting,burke2009feed,eckles2016estimating}}. However, we also \newrev{do not see strong evidence} of a backfire effect indicated by either \newrev{a consistently more negative effect of a cross-party interaction in predicting increased out-party activity, or a consistently more positive effect of a cross-party interaction in predicting increased in-party activity, compared with that of a same-party interaction}.
Our study sheds new light on the dynamics of political polarization on social media platforms by showing \rev{the nuanced after-effects} of cross-party communication in real-world settings.

\section{Related Work}
\spara{Political Polarization on Social Media}
\rev{While there is mixed evidence on whether social media exacerbates polarization in offline settings \cite{allcott2020welfare,asimovic2021testing}, online discussions on social media are shown to be extremely polarized.} Empirical studies of Twitter data show that Twitter \rev{retweet networks} are highly \rev{segregated} on various topics across multiple countries \cite{conover2011political,morales2015measuring,garimella2016quantifying,darwish2019quantifying,cossard2020falling, salloum2022separating}, \rev{indicating that retweeting -- usually a sign of endorsement \cite{metaxas2015retweets} -- is largely skewed toward the same party}. Similarly, researchers have found strong evidence of partisan sharing of information on Facebook \cite{an2014partisan,schmidt2018polarization}. 

\rev{The scarcity of cross-party endorsement hints at} the existence of echo chambers on social media, \rev{where users are exposed to similar views and isolated from opposing views. However,}
some suggest that the echo chamber narrative might be exaggerated \cite{barbera2020social,guess2018avoiding}.
For example, \citet{barbera2015tweeting} showed that cross-party interactions take place more frequently than believed on Twitter \rev{-- although these interactions can have a more negative tone \cite{mekacher2023systemic}}. \citet{barnidge2017exposure} found that more disagreement is perceived on social media than in other communication settings. \citet{fletcher2018people} revealed that social media users consumed more diverse news sources than non-users. \citet{flaxman2016filter} found the use of social networks and search engines to be associated with increased individual exposure to incongruent viewpoints, although it is also associated with increased ideological segregation between users. \rev{This challenges the argument that algorithms are what limit users' exposure to opposing views and thus exacerbate polarization.} \citet{bakshy2015exposure} further showed that individual selection plays a more important role than algorithms in limiting user exposure to cross-cutting content on Facebook\rev{, while \citet{robertson2023users} found that partisan news consumption on Google Search is driven primarily by user choices instead of algorithmic curation}.

Polarization dynamics also differ significantly across platforms. For example, \citet{cinelli2021echo} performed a comparative analysis of four well known social media platforms (Twitter, Facebook, Reddit, Gab) and found that Reddit embodied relatively smaller echo chamber effects than Facebook and Twitter in user interaction and information diffusion. Studies on Reddit more specifically showed the prevalence of cross-party interactions in Reddit political discussions \cite{defrancisci2021noecho}, and found more evidence of demographic instead of ideological segregation among user interactions on Reddit \cite{monti2023evidence}.

\spara{Effect of Cross-Party Interactions}
However, what is the effect of these cross-party interactions? Theory suggests that they may mitigate political polarization as intergroup contact reduces prejudice toward the outgroup \cite{allport1954nature,pettigrew1998intergroup}, but social experiments have shown mixed evidence regarding the effect of cross-party exposure or conversations. \citet{bail2018exposure} found that exposure to cross-party news on social media can increase ideological polarization, while others found scarce evidence of such a backfire effect (also referred to as backlash or boomerang effect) \cite{guess2020does,wojcieszak2023no,bakshy2015exposure}. A number of experimental studies on cross-party conversations suggest that they reduce affective polarization \cite{levendusky2021we,rossiter2020consequences} but not ideological polarization \cite{levy2021social,de2024cross}, while \citet{combs2023reducing} showed that anonymous cross-party conversations can decrease both affective and ideological polarization. Some researchers found that the depolarization effect takes place only when the conversation revolves around a casual topic \cite{santoro2022promise}, or when the interlocutors are not strong partisans \cite{thomsen2022intergroup}. Some further showed that negative cross-party contact can instead exacerbate out-party hostility \cite{wojcieszak2020can}.

Despite the considerable number of attempts to experimentally study the effect of cross-party interactions in synthetic settings, there has been limited empirical evidence on whether cross-party interactions depolarize real discussions on social media. Two survey studies before the rise of social media indicated that political diversity in one's social network may decrease one's participation in politics \cite{mutz2002consequences,huckfeldt2004disagreement}, and one empirical analysis of Facebook data echoes this finding by showing that political disagreement in one's Facebook network is associated with decreased turnout \cite{bond2015quantifying}. Most relevant to our work, \citet{beam2020facebook} showed that exposure to cross-cutting news is related to affective depolarization on Facebook, while \citet{heatherly2017filtering} found that engagement in cross-cutting discussions on social networking sites is associated with less affective polarization. However, both these studies used self-reported survey data and were not able to reveal user dynamics in unobserved settings.

\rev{Using real user behavior data, early studies have shown that social interaction and feedback in general boosts user activity in online communities \cite{joyce2006predicting,burke2009feed}, while this finding is confirmed by an experimental study \cite{eckles2016estimating}. However, the type of the interaction (i.e., cross-/same-party) presumably makes a substantial difference in its effect, as cross-party interactions are shown to be more negative \cite{marchal2022nice} and more toxic \cite{wu2021cross} than same-party ones on social media. In our work, we specifically study user dynamics following cross-party interactions.}

\section{Method}
We explore the effect of cross-party interactions by analyzing real user behavior data on Reddit. Reddit is a social media platform where users can create posts, comments, and replies of comments in different subreddits (i.e., subcommunities), while all the posts and comments can be voted up or down by other users.

We focus our analysis on the discussion of U.S. politics. On Reddit, the U.S. politics discussion is scattered across multiple subreddits, some of them being \textit{partisan subreddits} that are dominated by users of the same political leaning (e.g. \texttt{r/The\_Donald}, \texttt{r/hillaryclinton}), while some others are \textit{non-partisan subreddits} that engage users of different political leanings. A user's activity\footnote{Throughout this paper, we use ``activity'', ``participation'', or ``engagement'' to refer to the act of posting comments that receive a \textit{non-negative} score (i.e., number of upvotes subtracted by number of downvotes by other users) in a certain subreddit.} in partisan subreddits is often regarded an indicator of their political leaning \cite{defrancisci2021noecho}.

Considering the echo chamber narrative, we are interested in how users change their activity in partisan subreddits after a cross-party interaction in a non-partisan subreddit. Specifically, we raise the following research questions:
\begin{itemize}
    \item[RQ1] After receiving a cross-party reply in a non-partisan subreddit, do users \rev{increase out-party subreddit activity} (i.e., participate more in partisan subreddits of the opposite political leaning)?
    \item[RQ2] After receiving a cross-party reply in a non-partisan subreddit, do users \rev{increase in-party subreddit activity} (i.e., participate more in partisan subreddits of the same political leaning)?
\end{itemize}

\rev{While we are interested in the after-effects of a cross-party reply compared with the case of no reply, we also compare the after-effects of a cross-party reply with those of a same-party reply, in order to recognize the part that can be explained by the general motivating effect of social interaction \cite{joyce2006predicting,burke2009feed,eckles2016estimating}.} We detail our data and methods as follows.

\subsection{Data}
We use data from the Pushshift Reddit data set \cite{baumgartner2020pushshift}. 
We select \texttt{r/news} to be the non-partisan subreddit for sampling the cross-party interactions from, as it likely contains a large number of such interactions: consistently ranked among the top 15 most popular subreddits in previous years, \texttt{r/news} is considered the largest subreddit that discusses U.S. news. Moreover, echoing the idea that news discussion is one of the main loci where opinions are formed and challenged~\cite{kim1999news}, previous research identified how \texttt{r/news} accommodates both Democrats and Republicans, as well as interactions between them \cite{monti2023evidence}.

Considering the potential variation in dynamics across years, we collect data from five different years between 2014 and 2018. 
In order to have a feasible amount of data to analyze, for each year, we first collect all comments posted in \texttt{r/news} \rev{during a two-month \textit{observation period}. We select August to September as the observation period because it is relatively close to the Election Day in the U.S., and thus in election years we expect there to be a considerable amount of discussion related to U.S. politics during this period. In the meantime, the period is not so close to the Election Day that the election results will play a major role in the discussion dynamics.} Statistics of the collected data are reported in Table~\ref{tab:stats}. 

We extract the set of users who were active in this period, removing deleted accounts and automated accounts as detected by a previous Reddit study \cite{rollo2022communities}. \rev{For each user, we pinpoint the time of their first received reply in \texttt{r/news} during the observation period; for those who did not receive any reply, we take the time of their first posted comment in \texttt{r/news} during the observation period. We consider 30 days before that as the \textit{before period} of this user, and 30 days after as the \textit{after period}. Then, we collect all comments they posted in \emph{any} subreddit in the before period and the after period.}
As we will describe in the following, we use this information for characterizing the political leaning of each user, and for evaluating whether they increased their activity in partisan subreddits after an interaction in the observation period.

\begin{table*}[htb]
    \centering
    \resizebox{\textwidth}{!}{
    \begin{tabular}{lrrrrr}
    \toprule
     & 2014 & 2015 & 2016 & 2017 & 2018 \\
    \midrule
    \# comments & 822,237 & 1,156,138 & 1,154,495 & 1,621,039 & 1,424,752 \\
    \# interactions & 456,204 & 749,276 & 760,070 & 1,123,115 & 897,844 \\
    \# accounts & 121,886 & 164,292 & 163,798 & 211,307 & 220,338 \\
    \# users & 121,571 & 163,874 & 163,256 & 210,522 & 219,509 \\
    \# non-zero-leaning users (left / right) & 15,705 / 11,406 & 24,931 / 16,005 & 34,610 / 23,462 & 51,691 / 26,832 & 58,830 / 22,584 \\
    \midrule
    \# users receiving reply & 8,068 & 13,230 & 22,512 & 31,651 & 29,923 \\
    \# users receiving sociopolitical reply & 3,768 & 5,998 & 10,283 & 15,135 & 14,024 \\
    \# users receiving cross-party reply (left / right) & 917 / 941 & 1,407 / 1,434 & 2,316 / 2,551 & 3,811 / 3,711 & 3,177 / 3,204 \\
    \# users receiving same-party reply (left / right) & 1,292 / 618 & 2,281 / 876 & 4,016 / 1,400 & 6,002 / 1,611 & 6,585 / 1,058 \\
    \# users receiving no reply (left / right) & 2,250 / 1,849 & 3,315 / 2,300 & 4,852 / 3,867 & 7,717 / 4,190 & 9,486 / 3,608 \\
    \bottomrule
    \end{tabular}
    }
    \caption{Statistics of the data set. The upper half shows the total number of comments in the observation period, the number of comment-reply interactions, the total number of accounts involved, the number of users excluding deleted and automated accounts, and the number of users with a \rev{non-zero} inferred leaning. The lower half shows \rev{the statistics of our regression data set after filtering, including the number of users receiving a valid reply (i.e., both the user and the replier have a non-zero inferred leaning), those receiving a sociopolitical reply, those receiving a sociopolitical cross-party reply, those receiving a sociopolitical same-party reply, and those who received no reply during the observation period. For four of the rows, we report separately the number of users who are inferred to be left-/right-leaning.}}
    \label{tab:stats}
\end{table*}

\subsection{Leaning Detection}
To identify cross-party interactions and partisan subreddit activity, we need to evaluate the leaning of each user and subreddit. 

For this purpose, we first identify a list of \textit{political subreddits}.
To do so, we manually inspect for each year the top 1000 subreddits with the most comments in our data: for each subreddit, we scan through its first 5 posts and label it as a political subreddit if there is at least one post related to U.S. politics. With this approach, we find 87 political subreddits in total.

Then, we determine the leaning of each subreddit based on its political leaning score as given by \citet{waller2021quantifying}. \rev{In their paper, they devised a method for inferring the position of each subreddit along different social dimensions, by comparing the user base of the subreddit with that of certain seed subreddits with known positions (e.g., \texttt{r/democrats} and \texttt{r/Conservative} along the political dimension). Note that using this approach, the leaning score of each subreddit essentially reflects the leaning of its user base, as opposed to that of its topic.}

We use a binned version of the subreddit leaning scores as done in the original paper: in this way, each subreddit $s$ is associated with a political leaning score $\ell_s \in \{-2,-1,0,1,2\}$, where -2 indicates \rev{extremely} left-wing, -1 indicates \rev{left-wing}, 0 indicates non-partisan, 1 indicates \rev{right-wing}, and 2 indicates \rev{extremely} right-wing. The list of political subreddits and their leaning scores can be found in the Appendix.

For each user $u$, we infer their leaning $\ell_u$ based on their activity in political subreddits during the \textit{before} period.
Specifically, we define an activity vector $p_u$ for each user, where for each political subreddit $s$, \rev{we have its corresponding entry $p_{u,s}$ as the ratio of comments that user $u$ posted in subreddit $s$ over the total number of comments user $u$ posted in all political subreddits during the before period}. 
Note that we only count comments that received non-negative scores as valid activity\rev{: in this way, we exclude from the leaning calculation those comments that are negatively received by other users in the subreddit, which could indicate negative behavior such as attacking or trolling, as opposed to proper engagement with the subreddit.}

Then, the political leaning score $\ell_u$ of a user $u$ is obtained as the weighted average of the subreddits leaning, i.e. $\ell_u = \sum_{s} p_{u,s} \cdot \ell_s$. We assign a democratic leaning to a user if their leaning score is below 0, and a republican leaning if their leaning score is above 0. 

A \emph{cross-party interaction} is then defined as a comment-reply interaction between two users who have opposite leanings, \rev{while a \emph{same-party interaction} is a comment-reply interaction between two users of the same leaning}. Similarly, \rev{out}-party (resp. \rev{in}-party) subreddit activity is defined as a user participating in a subreddit of the opposite (resp. same) leaning. 

\rev{We create a regression data set where each row corresponds to one user and their first comment-reply interaction (if any) in \texttt{r/news} during the observation period.} To ensure that both cross-/same-party interactions and out-/in-party subreddit activity are well-defined, we select the rows where \rev{the user has a non-zero inferred leaning; and for rows with an interaction, we selected those where} both the user (i.e., comment author) and the replier (i.e., reply author) have non-zero inferred leanings.

\rev{
To exclude irrelevant discussions, we limit the scope of our analysis to \emph{sociopolitical} interactions. We consider a comment sociopolitical if it discusses at least one of the following:
(i)~political figures, parties, or institutions;
(ii)~broad cultural and social issues (e.g., civil rights, moral values);
(iii)~national issues (e.g., healthcare, welfare),
following the definition by \citet{moy2006predicting}. To select only comment-reply interactions where the reply is sociopolitical, we use the classifier developed by~\citet{monti2022language} that was trained on a large Reddit data set and obtained an F1-score of 82\% on a manually labeled subset.} Statistics of the regression data set after filtering are presented in Table~\ref{tab:stats}.

\begin{figure*}[htbp]
\centering
\includegraphics[width=0.85\textwidth]{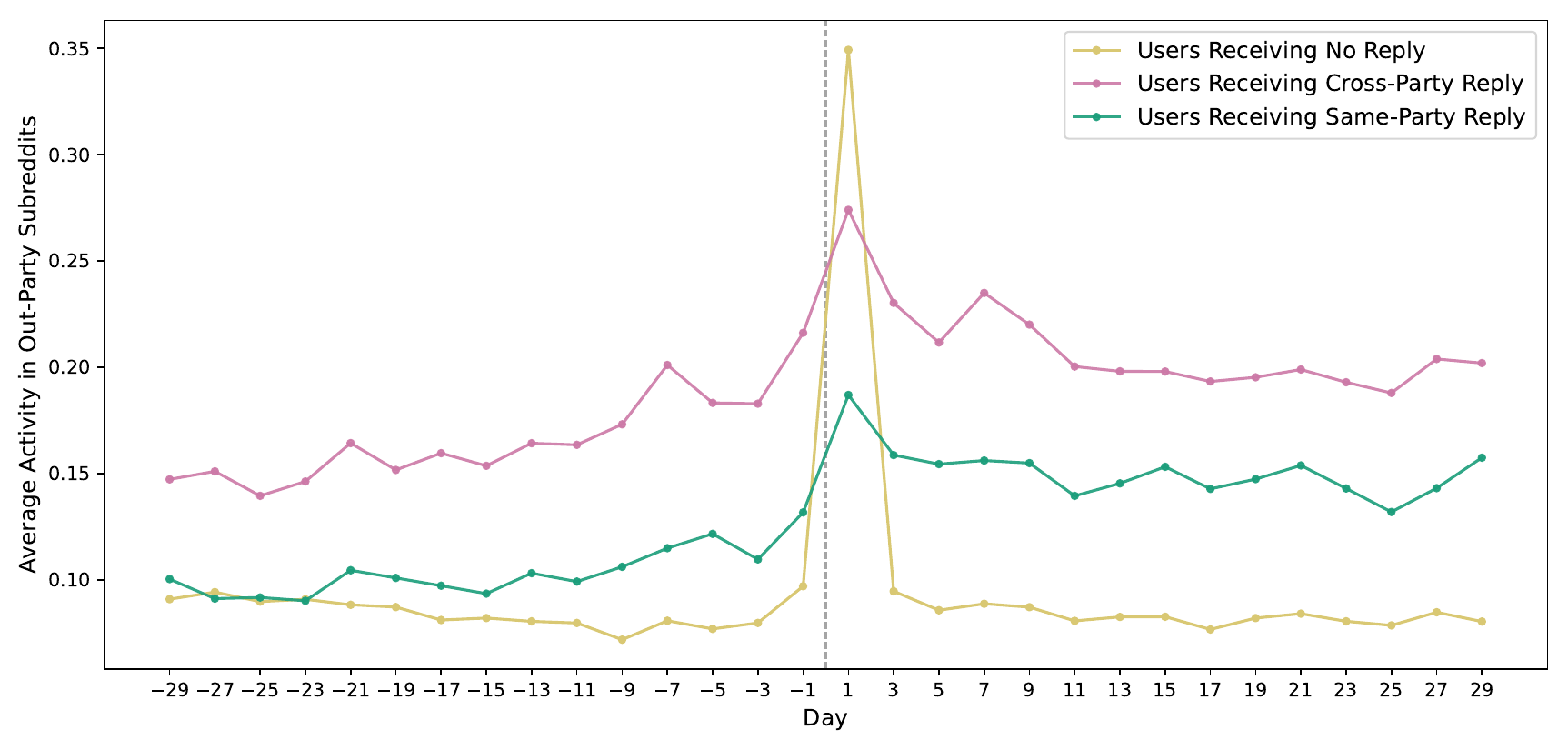}
\caption{\rev{Average user activity in out-party subreddits for every two-day interval, from 30 days before to 30 days after receiving the first reply (resp. posting the first comment) in \texttt{r/news} in the observation period, for users who received replies (resp. users who received no replies).}
}
\label{fig:actv}
\end{figure*}

\begin{table*}[t]
    \centering
    \setlength{\tabcolsep}{10pt}
    \renewcommand{\arraystretch}{1.05}
    \resizebox{0.95\textwidth}{!}{
    \begin{tabular}{llllll}
\toprule
{} &          A &          B &          C &          D &          E \\
\midrule
$\Delta$ BIC                            &          0 &        -76 &        -30 &        -65 &        -55 \\
\midrule

Cross-Party Interaction (Top-Level)        &            &      0.080 &      0.057 &      0.080 &      0.078 \\
Cross-Party Interaction (Nested) &            &   0.177*** &   0.156*** &   0.177*** &   0.177*** \\
Same-Party Interaction (Top-Level)         &            &   0.189*** &    0.164** &   0.188*** &    0.157** \\
Same-Party Interaction (Nested)  &            &   0.262*** &   0.242*** &   0.262*** &   0.262*** \\ [4pt]

Left-Wing User                 &  -0.686*** &  -0.688*** &  -0.688*** &  -0.688*** &  -0.688*** \\
Hyper-Partisan User            &  -0.318*** &  -0.241*** &  -0.241*** &  -0.241*** &  -0.241*** \\
User Out-Party Activity        &   0.208*** &   0.213*** &   0.213*** &   0.213*** &   0.213*** \\
User In-Party Activity         &   0.115*** &   0.108*** &   0.108*** &   0.108*** &   0.108*** \\
User Non-Partisan Activity     &   0.259*** &   0.189*** &   0.189*** &   0.189*** &   0.189*** \\ [4pt]

Reply Length                   &            &            &     -0.021 &            &            \\
Reply with URL                 &            &            &     -0.027 &            &            \\
Negative Reply Text            &            &            &      0.048 &            &            \\
Positive Reply Text            &            &            &     -0.014 &            &            \\
Toxic Reply Text               &            &            &      0.079 &            &            \\ [4pt]

Reply Score                    &            &            &            &      0.001 &            \\ [4pt]

Cross-Party Conflictual Scores &            &            &            &            &      0.006 \\
Same-Party Conflictual Scores  &            &            &            &            &      0.106 \\

\bottomrule
\end{tabular}
    }
    \caption{Results of the model selection test for choosing the most important independent variables, where we compare different regression models using aggregated data from all considered years.
    \rev{Each column (A, B, ...) corresponds to a regression model with a different set of independent variables, where the dependent variable is \emph{More Out-Party Activity}.}
    $\Delta$ BIC indicates the BIC score of the model compared with the baseline model $A$, where a lower BIC value indicates a better-performing model.
    Significance of each variable marked by ***: $p<0.001$, **: $p<0.01$, *: $p<0.05$.
    }
    \label{tab:model-comparison}
\end{table*}

\subsection{Average User Activity}
\rev{To gain an initial understanding of user activity patterns before and after an interaction, we plot the average user activity in out-party subreddits for every two-day interval, from 30 days before the interaction to 30 days after (Figure~\ref{fig:actv}), respectively for users receiving a cross-party reply and for those receiving a same-party reply.}

\rev{We see that on average, out-party activity seems to have already increased before the interaction, presumably because of an activity peak around the time of the interaction. To provide a baseline of peaked activity, we additionally plot the average out-party activity of users receiving no reply, from 30 days before their first posted comment in \texttt{r/news} during the observation period to 30 days after. We find that compared with the baseline, the increase in out-party activity after an interaction seems to sustain for much longer, suggesting the existence of post-interaction dynamics that are not explained by peaked activity.}

\subsection{Regression Analysis}
\rev{While the activity plot already reveals interesting patterns, it only presents compressed information from averaged statistics. To utilize individual-level data when controlling for potential confounding factors, we perform a regression analysis as described below.}

In order to evaluate whether users increase their activity in partisan subreddits after receiving a reply in \texttt{r/news}, we consider two dependent variables for our regression model:
\begin{itemize}
    \item \textit{More Out-Party Activity}: 1 if the user had more out-party subreddit activity in the after period than in the before period, and 0 otherwise.
    \item \textit{More In-Party Activity}: 1 if the user had more in-party subreddit activity in the after period than in the before period, and 0 otherwise.
\end{itemize}

\rev{The key of the regression analysis is to compare the activity of users who received a cross-/same-party reply and the activity of those who did not receive any reply (i.e., baseline case), while controlling for other user features.}
Naturally, we have \textit{Cross-Party Interaction} and \textit{Same-Party Interaction} as the main independent variables in our regression model. \rev{Further, we speculate that there may exist a substantial difference between the comment-reply interactions where the comment is a top-level reply to some post (abbreviated as ``top-level''), and those where the comment is a reply to some other comment (abbreviated as ``nested''). In the latter case, especially if the comment replies to an out-party member, it is more likely that the user (i.e., comment author) already possesses a motivation to interact with the out-party, and thus may also increase activity in out-party subreddits. Therefore, we include four separate interaction variables in our regression model:} 

\begin{itemize}
    \item \rev{\textit{Cross-Party Interaction (Top-Level)}: 1 if the user received a cross-party reply to a comment that is a top-level reply to some post, and 0 otherwise.}
    \item \rev{\textit{Cross-Party Interaction (Nested)}: 1 if the user received a cross-party reply to a comment that is a reply to some other comment, and 0 otherwise.}
    \item \rev{\textit{Same-Party Interaction (Top-Level)}: 1 if the user received a same-party reply to a comment that is a top-level reply to some post, and 0 otherwise.}
    \item \rev{\textit{Same-Party Interaction (Nested)}: 1 if the user received a same-party reply to a comment that is a reply to some other comment, and 0 otherwise.}
\end{itemize}

Meanwhile, we introduce the following variables to control for the characteristics of the user as well as other features of the reply:

\begin{itemize}
    \item User leaning:
    \begin{itemize}
        \item \textit{Left-Wing User}: 1 if the user has a leaning score below 0, and 0 otherwise.
        \item \textit{Hyper-Partisan User}: 1 if the user has a leaning score of -2 or 2, and 0 otherwise.
    \end{itemize}
\end{itemize}

\begin{itemize}
    \item User activity:
    \begin{itemize}
        \item \textit{User Out-Party Activity}: Number of comments posted by the user in subreddits of the opposite leaning in the before period, in logarithmic form\rev{, standardized by subtracting the mean and dividing by 2 standard deviations (an approach suggested by \citet{gelman2008scaling})}.
        \item \textit{User In-Party Activity}: Number of comments posted by the user in subreddits of the same leaning in the before period, in logarithmic form, standardized by subtracting the mean and dividing by 2 standard deviations.
        \item \textit{User Non-Partisan Activity}: Number of comments posted by the user in non-partisan subreddits in the before period, in logarithmic form, standardized by subtracting the mean and dividing by 2 standard deviations.
    \end{itemize}
\end{itemize}

\begin{itemize}
    \item Reply text (only for users who received a reply):
    \begin{itemize}
        \item \textit{Reply Length}: Number of characters in the reply, in logarithmic form, standardized by subtracting the mean and dividing by 2 standard deviations.
        \item \textit{Reply with URL}: 1 if the reply contains URL links, and 0 otherwise.
        \item \textit{Negative Reply Text}: 1 if the VADER sentiment score \cite{hutto2014vader} of the reply text is below 0, and 0 otherwise. The score falls in the range of $[-1, 1]$, where -1 indicates the most negative sentiment, 1 indicates the most positive sentiment, and 0 indicates neutral sentiment.
        \item \textit{Positive Reply Text}: 1 if the VADER sentiment score \cite{hutto2014vader} of the reply text is above 0, and 0 otherwise.
        \item \textit{Toxic Reply Text}: 1 of the toxicity score of the reply text returned by the Perspective API\footnote{\url{https://perspectiveapi.com}} is above 0.7, and 0 otherwise. The score falls in the range of $[0, 1]$, where 0 indicates the least toxic, and 1 indicates the most toxic.
    \end{itemize}
\end{itemize}

\begin{itemize}
    \item Reply score (only for users who received a reply):
    \begin{itemize}
        \item \textit{Reply Score}: The user-voted score of the reply in logarithmic form, standardized by subtracting the mean and dividing by 2 standard deviations.
        \item \rev{\textit{Cross-Party Conflictual Scores}: 1 if the user and the replier have opposite leanings while} the comment and the reply have scores of opposite signs, and 0 otherwise. Scores of opposite signs have been used in previous research as an indicator of a conflict between the parent author and the child author~\cite{monti2020learning}.
        \item \rev{\textit{Same-Party Conflictual Scores}: 1 if the user and the replier have the same leaning while} the comment and the reply have scores of opposite signs, and 0 otherwise. 
    \end{itemize}
\end{itemize}

\section{Results}
In this section, we first present the results of a model selection test, where we use the data aggregated from all studied years to select the most important independent variables for predicting the dependent variable. Following that, we report the results of the regression analysis where we run a separate regression model on each year's data, with the aim of revealing time-specific dynamics as well as dynamics that are consistent throughout time.

\medskip

\spara{Model selection}
To select the most important independent variables, we compare regression models with different sets of variables. Specifically, we gradually add groups of variables onto a baseline regression model and compare the Bayesian information criterion (BIC) \cite{schwarz1978estimating} values of the models before and after adding each group of variables. Since a lower BIC indicates a better model, we retain the group of variables if the BIC of the model decreases after adding the variables. Table~\ref{tab:model-comparison} shows the models we tested and their difference from the baseline model in BIC, where the dependent variable is \textit{More Out-Party Activity} for all models. 

We start with baseline model $A$ that considers all the user leaning and user activity variables.
Upon adding the \rev{interaction variables}, the BIC of the model decreases by 76 (model $B$), which indicates that the \rev{interaction variables substantially increase} the model explanatory power even when considering a penalty term for the increased number of variables. Therefore, we include the interaction variables in all following models. 

By contrast, upon adding the reply text variables (model $C$), the model BIC increases, which means that these variables degrade the predictive power of the model, and thus we exclude them from our analysis.
\rev{Similarly, none of the reply score variables (models $D$ and $E$) is shown to be important to the model}. Therefore, we choose model \rev{$B$ (with only interaction, user leaning, and user activity variables)} as the regression model for running our yearly analysis.

\rev{Note that we have also tried running the variable selection process on the other part of the data where the reply of the interaction is classified as non-sociopolitical. There, including the interaction variables increases the BIC of the model by 6, which indicates a decrease in its predictive power. This suggests that the correlations we find between interactions and subreddit activity in the sociopolitical counterpart of the data likely indicate real dynamics in political discussions, as opposed to artifacts of Reddit data that are present in both sociopolitical and non-sociopolitical interactions.}

\begin{figure*}[htbp]
\centering
\includegraphics[width=\textwidth]{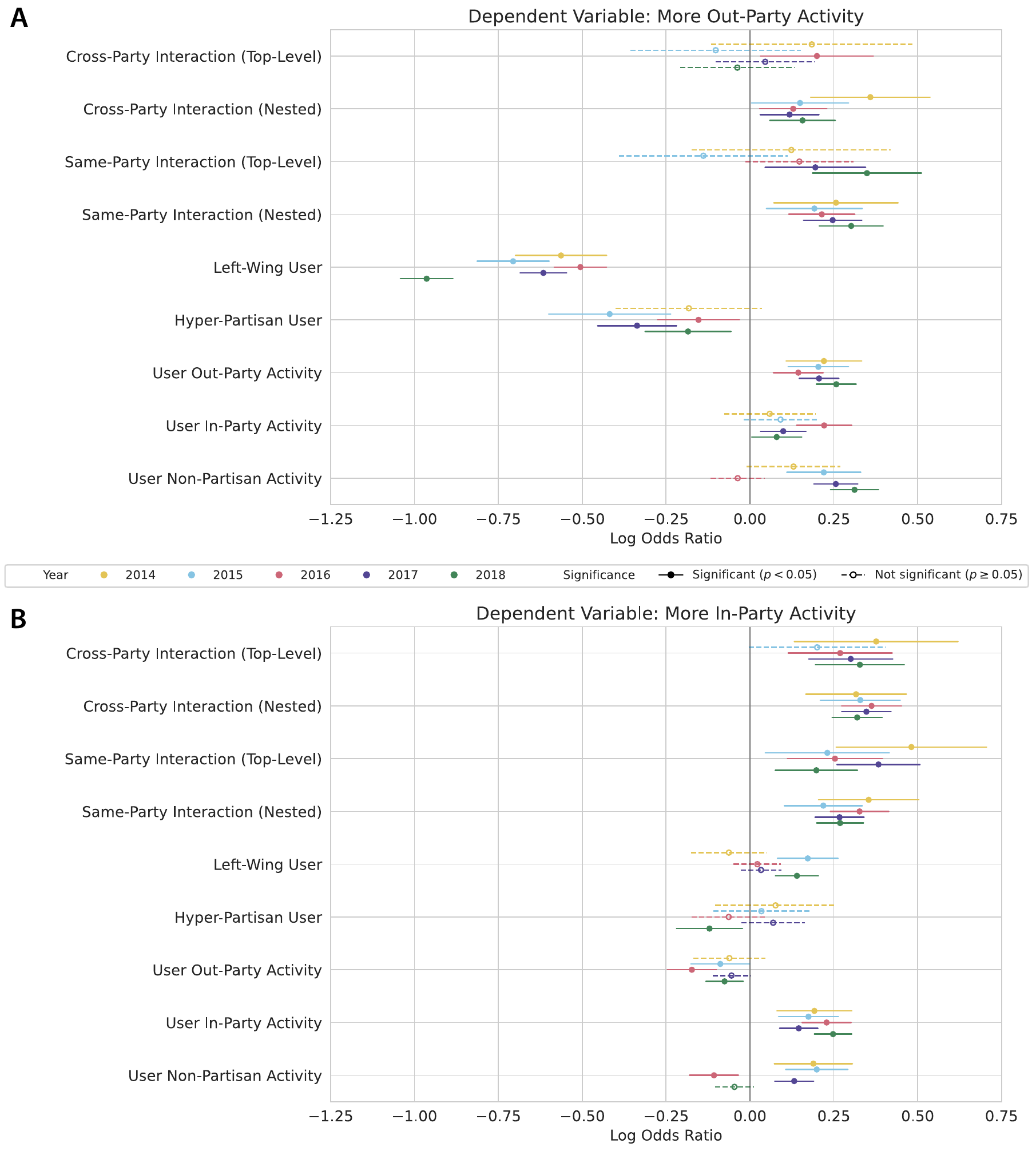}
\caption{Coefficients (i.e., log odds ratio) of the regression model for predicting \rev{A) increased out-party subreddit activity, and B) increased in-party subreddit activity}.
Each row corresponds to an independent variable. Each point on the row corresponds to the estimated coefficient of the independent variable in one year's model\rev{, and the bar underneath indicates the 95\% confidence interval of the estimate}. A solid circle \rev{with a solid line} indicates that the independent variable has a significant correlation with the dependent variable in the corresponding year, while a hollow circle \rev{with a dotted line} indicates a non-significant one. A coefficient above 0 indicates that the independent variable has a positive correlation with the dependent variable, while a coefficient below 0 indicates a negative one. 
}
\label{fig:res}
\end{figure*}

\spara{Aggregated analysis}
\rev{Across all models with interaction variables, we first see that receiving a nested cross-party interaction (i.e., receiving a reply to a comment that is already a reply to some other comment) seems significantly associated with more activity in out-party subreddits after the interaction ($p<0.001$), while receiving a top-level cross-party interaction (i.e., receiving a reply to a comment that is a top-level reply to some post) demonstrates no significant association with more out-party activity afterward. This is in line with our speculation that users who receive a nested cross-party interaction might already possess a higher level of motivation to interact with the out-party, and thus are more likely to engage with out-party subreddits after the interaction. With this confounding effect excluded, our results indicate that a cross-party interaction alone is not significantly associated with an increase in subsequent out-party activity.}

\rev{Interestingly, we observe that receiving a same-party interaction, top-level or nested, is significantly associated with more activity in out-party subreddits after the interaction ($p<0.001$ for both in the selected model $B$), while the coefficients are consistently larger than those in the corresponding cross-party case. We will further corroborate and analyze these results in the yearly analysis.}

We also find that in all tested models, left-wing users are less likely to increase their activity in right-wing subreddits than vice versa ($p < 0.001$).
This result aligns with previous findings on how right-wing users tend to be more involved in cross-cutting interactions than left-wing users \cite{wu2021cross,eady2019many,grinberg2019fake}. 
Similarly, we find that hyper-partisan users who dominantly participate in hyper-partisan subreddits are less likely to increase activity in out-party subreddits ($p < 0.001$), echoing previous studies that showed how hyper-partisan individuals are more difficult to depolarize \cite{wojcieszak2011deliberation,thomsen2022intergroup,de2024cross}.

\medskip

\spara{Yearly analysis}
We then run regression model \rev{$B$} separately on each year between 2014 and 2018. The coefficients of each year's regression model that predicts \textit{More Out-Party Activity} are plotted in Figure~\ref{fig:res}A.
In order to answer RQ2, we run the same regression model to predict the 
dependent variable \textit{More In-Party Activity}, the coefficients of which are plotted in Figure~\ref{fig:res}B.

We see that \rev{both nested interaction variables have significant positive correlations with both dependent variables in all years. This means that receiving a nested reply in \texttt{r/news}, whether cross-party or same-party, is consistently associated with a user's increased activity in both out-party and in-party subreddits. This can be explained by how posting a comment as a reply to some other comment already reflects the user's elevated intention to actively engage with other opinions, which helps boost their subreddit activity when they receive a reply to the comment.} 

The coefficients of the top-level interaction variables demonstrate the correlation between receiving a reply and subreddit activity without the confounding effect in nested conversations. Figure~\ref{fig:res}A shows that receiving a reply to a top-level comment in \texttt{r/news}, whether cross-party or same-party, does not have a consistently significant association with increased out-party activity. However, the association is positive and significant for cross-party interaction in the election year 2016, and for same-party interaction in year 2017 and 2018, which might reflect some special discussion dynamics during the election period as well as after the new president took office. In general, the results do not indicate a strong depolarization effect following cross-party interactions, but we also do not see strong evidence of a backfire effect, in which case a cross-party interaction is expected to be more negatively associated with increased out-party activity \newrev{than a same-party interaction: Figure~\ref{fig:res-diff}A shows that the association is not significantly more negative across all years (see Appendix for a more detailed discussion).}

\rev{Meanwhile, in five (resp. four) out of five years, receiving a same-party (resp. cross-party) reply to a top-level comment in \texttt{r/news} is significantly associated with increased in-party subreddit activity (Figure~\ref{fig:res}B).} This can be explained by the general motivating effect of social interaction in online communities \cite{joyce2006predicting,burke2009feed,eckles2016estimating}. Here we do not observe a significantly larger effect of a cross-party interaction than that of a same-party interaction \newrev{(Figure~\ref{fig:res-diff}B)}, again finding no evidence of a backfire effect, in which case a cross-party interaction is expected to be more positively associated with increased in-party activity.

\rev{It is also interesting that the difference between the effect of a top-level interaction and that of a nested interaction is more pronounced in predicting increased out-party activity than in predicting increased in-party activity. We speculate that it is because out-party community engagement requires a much stronger motivation, as there exist substantial barriers toward intergroup contact \cite{ron2017willingness}. Therefore, the type of the interaction, which reflects the user's initial level of willingness to engage with other opinions, becomes a more important predictor of increased out-party activity.}

Among other variables, we find confirmation that left-wing users are significantly less likely to increase activity in out-party subreddits (Figure~\ref{fig:res}A). Notably, the effect size of the \textit{Left-Wing User} variable in predicting increased out-party activity is substantially larger than those of the interaction variables, indicating that user leaning probably plays a more critical role than social interactions in predicting out-party community engagement. Consistent with the results of our aggregated data analysis, hyper-partisan users are also significantly less likely to participate more in out-party subreddits (Figure~\ref{fig:res}A). \rev{However, user leaning seems to be a weak predictor of increased in-party activity.}

\newrev{In addition to our main regression model, we have also run an alternative model where the dependent variable is increased ratio of out-party subreddit activity. The results mostly echo our main findings (see Appendix for a more detailed discussion).}

\section{Discussion}
\spara{Findings \& Implications}
We find that in the U.S. politics discussion on Reddit, receiving a cross-party reply to a posted comment in \texttt{r/news} is not necessarily associated with increased subsequent activity in out-party subreddits, unless the comment itself is already a reply to some other comment (RQ1). This indicates that the depolarization effect of a cross-party interaction in boosting engagement with out-party communities is \newrev{highly conditioned on the receiver's intention to actively engage with the opinions of others. Even when there is a depolarization effect, it likely reflects a more general dynamic of feedback-boosted engagement that also explains a similar association between a same-party interaction and increased activity in out-party communities.}

\rev{On the other hand, we do not see strong evidence that a cross-party interaction in a non-partisan community comes with a backfire effect of decreasing out-party activity or increasing in-party activity\newrev{, beyond what can be explained by feedback-boosted engagement (RQ2)}. Compared with the predictive effect of a same-party interaction in our regression models, the effect of a cross-party interaction in predicting increased out-party activity is not consistently more negative, while the effect of a cross-party interaction in predicting increased in-party activity is not consistently more positive. This finding aligns with previous studies that found scarce backfire effects of exposure to counter-partisan information \cite{guess2020does,wojcieszak2023no,bakshy2015exposure}}.

\spara{Limitations}
Our study used a limited data sample for the analysis. We ran the analysis on a single social media platform (i.e., Reddit), on a single community of interaction (i.e., \texttt{r/news}), \rev{within a data collection period of four months (i.e., July to October)} across five different years (i.e., 2014-2018), \rev{and specifically in the U.S. politics context}. While we look forward to future studies that replicate our findings under other settings, the results of our study are not guaranteed to generalize to other social media platforms, communities of interaction, time frames, \rev{or contexts}. \rev{Also, we studied the effect following a single interaction, where multiple might be needed toward a more substantial change in user behavior.}

For conducting a quantitative analysis, we had to operationalize theoretical constructs using observable user behavior logs on Reddit with some level of abstraction, and this process could have introduced inaccuracy into our conclusions. \rev{The user leaning inference in particular could have been disturbed by multiple sources of noise: we inferred a user's leaning based on their activity in partisan subreddits, where the partisan subreddits were identified based on a leaning score of each subreddit that only reflects the average leaning of its user base, and subreddit activity is measured by the number of non-negative-score comments posted; we also worked with the assumption that users dominantly engage with in-party subreddits, which may not always hold.} While we hope our current approach provides a good approximation for the corresponding theoretical constructs, our conclusions should be interpreted with this limitation in mind. 
\rev{More precise methods for detecting the political leaning of users, for example those that extract information also from the text they post, will help validate our results.}

Finally, our study uses a regression analysis approach, \rev{which processes time series data in aggregated form. While such approach reveals user dynamics on a broader scale, future studies can deploy event history and time series analysis methods that make use of more fine-grained information. Moreover, regression analysis} uncovers correlation instead of causality, therefore the effect we observed between cross-party interactions and subreddit activity could have been confounded by other factors. As much as we tried to control for confounding variables that are observable on Reddit (e.g., \rev{top-level or nested interaction,} user leaning, user activity), we were not able to rule out the effect of unobservable confounders (e.g., a user's increased interest in the opposite party that stems from their real-life interactions might increase both their likelihood of receiving a cross-party reply in \texttt{r/news} and their likelihood of increasing activity in out-party subreddits). Further evidence is needed to establish a causal link for the correlation we found.
However we believe that our work, by presenting a correlation analysis, serves as an important first step toward casual studies that further answer the questions we have.

\section*{Acknowledgments}
We thank the anonymous reviewers for their insightful comments on our manuscript. In addition, MK acknowledges funding from the Research Council of Finland (349366; 352561).

\bibliography{references}

\clearpage

\section{Ethics Checklist}
\begin{enumerate}

\item For most authors...
\begin{enumerate}
    \item  Would answering this research question advance science without violating social contracts, such as violating privacy norms, perpetuating unfair profiling, exacerbating the socio-economic divide, or implying disrespect to societies or cultures?
    \answerYes{Yes, see the Ethics Statement section.}
  \item Do your main claims in the abstract and introduction accurately reflect the paper's contributions and scope?
    \answerYes{Yes.}
   \item Do you clarify how the proposed methodological approach is appropriate for the claims made? 
    \answerYes{Yes.}
   \item Do you clarify what are possible artifacts in the data used, given population-specific distributions?
    \answerYes{Yes, see the Limitations subsection in the Discussion section.}
  \item Did you describe the limitations of your work?
    \answerYes{Yes, see the Limitations subsection in the Discussion section.}
  \item Did you discuss any potential negative societal impacts of your work?
    \answerYes{Yes, see the Ethics Statement section below.}
      \item Did you discuss any potential misuse of your work?
    \answerYes{Yes, see the Ethics Statement section below.}
    \item Did you describe steps taken to prevent or mitigate potential negative outcomes of the research, such as data and model documentation, data anonymization, responsible release, access control, and the reproducibility of findings?
    \answerYes{Yes, see the Ethics Statement section below.}
  \item Have you read the ethics review guidelines and ensured that your paper conforms to them?
    \answerYes{Yes.}
\end{enumerate}

\item Additionally, if your study involves hypotheses testing...
\begin{enumerate}
  \item Did you clearly state the assumptions underlying all theoretical results?
    \answerNA{NA}
  \item Have you provided justifications for all theoretical results?
    \answerNA{NA}
  \item Did you discuss competing hypotheses or theories that might challenge or complement your theoretical results?
    \answerNA{NA}
  \item Have you considered alternative mechanisms or explanations that might account for the same outcomes observed in your study?
    \answerNA{NA}
  \item Did you address potential biases or limitations in your theoretical framework?
    \answerNA{NA}
  \item Have you related your theoretical results to the existing literature in social science?
    \answerNA{NA}
  \item Did you discuss the implications of your theoretical results for policy, practice, or further research in the social science domain?
    \answerNA{NA}
\end{enumerate}

\item Additionally, if you are including theoretical proofs...
\begin{enumerate}
  \item Did you state the full set of assumptions of all theoretical results?
    \answerNA{NA}
	\item Did you include complete proofs of all theoretical results?
    \answerNA{NA}
\end{enumerate}

\item Additionally, if you ran machine learning experiments...
\begin{enumerate}
  \item Did you include the code, data, and instructions needed to reproduce the main experimental results (either in the supplemental material or as a URL)?
    \answerNA{NA}
  \item Did you specify all the training details (e.g., data splits, hyperparameters, how they were chosen)?
    \answerNA{NA}
     \item Did you report error bars (e.g., with respect to the random seed after running experiments multiple times)?
    \answerNA{NA}
	\item Did you include the total amount of compute and the type of resources used (e.g., type of GPUs, internal cluster, or cloud provider)?
    \answerNA{NA}
     \item Do you justify how the proposed evaluation is sufficient and appropriate to the claims made? 
    \answerNA{NA}
     \item Do you discuss what is ``the cost`` of misclassification and fault (in)tolerance?
    \answerNA{NA}
  
\end{enumerate}

\item Additionally, if you are using existing assets (e.g., code, data, models) or curating/releasing new assets, \textbf{without compromising anonymity}...
\begin{enumerate}
  \item If your work uses existing assets, did you cite the creators?
    \answerYes{Yes, see the Data subsection in the Method section.}
  \item Did you mention the license of the assets?
    \answerYes{Yes, see the Ethics Statement section below.}
  \item Did you include any new assets in the supplemental material or as a URL?
    \answerNo{No, we only publish our code and preprocessed data in a github repository (URL anonymized now for review).}
  \item Did you discuss whether and how consent was obtained from people whose data you're using/curating?
    \answerYes{Yes, see the Ethics Statement section below.}
  \item Did you discuss whether the data you are using/curating contains personally identifiable information or offensive content?
    \answerYes{Yes, see the Ethics Statement section below.}
\item If you are curating or releasing new datasets, did you discuss how you intend to make your datasets FAIR?
    \answerNA{NA}
\item If you are curating or releasing new datasets, did you create a Datasheet for the Dataset? 
    \answerNA{NA}
\end{enumerate}

\item Additionally, if you used crowdsourcing or conducted research with human subjects, \textbf{without compromising anonymity}...
\begin{enumerate}
  \item Did you include the full text of instructions given to participants and screenshots?
    \answerNA{NA}
  \item Did you describe any potential participant risks, with mentions of Institutional Review Board (IRB) approvals?
    \answerNA{NA}
  \item Did you include the estimated hourly wage paid to participants and the total amount spent on participant compensation?
    \answerNA{NA}
   \item Did you discuss how data is stored, shared, and deidentified?
   \answerNA{NA}
\end{enumerate}
\end{enumerate}

\section*{Ethics Statement}
We collected the publicly available Reddit data when the Reddit Data API granted free access as well as a worldwide, non-exclusive, non-transferable, non-sublicensable, and revocable license to all users \cite{redditapiterms}. Participants who generated the data agreed to the Reddit Privacy Policy \cite{redditprivacypolicy}, and were aware of the public and freely accessible nature of the content they posted, as the subreddits are public, not password-protected, and are participated by thousands of active subscribers. Reddit's use of pseudonymous accounts reduces the likelihood of uncovering the true identity of users, and we removed those pseudonyms in our preprocessing as a precaution. Importantly, our findings offer only aggregated estimates without disclosing individual details.

To make our study reproducible, we host the preprocessed data and the code used for our analysis in a public Github repository\footnote{\url{https://github.com/yanxxia/cross-party-reddit}}. While the original data may contain personally identifiable information or offensive content, in our preprocessed data, we have removed all user identifiers and converted the content of comments into linguistic feature scores.

There are potential broader impacts of our research. On the positive side, understanding the relationship between user interactions, user engagement, and political polarization in online discussions can inform the design of healthier social media platforms that alleviate polarization among users.
However, we acknowledge that our conclusions can be misinterpreted in an overly simplified and generalized way (e.g., ``cross-party interactions on Reddit decrease political polarization'', ``the content of a reply does not change the effect it brings'', etc.), potentially fueling inappropriate behavior in online discussions. Here we would like to reiterate that our conclusions are limited by the technical approach we have taken for our analysis, and that they should always be interpreted within the context of our study.

\appendix
\section{Appendix A: Additional Regression Results}
\label{sec:add-res}
\spara{Difference-in-effects models}
\newrev{In addition to our main regression model that shows the absolute effects of cross-party interactions and same-party interactions, we fit another \emph{difference-in-effects} model to inspect the differences in their effects. In this alternative model, we replace the \emph{Same-Party Interaction} variables with \emph{Interaction} variables, so that the coefficient of a \emph{Cross-Party Interaction} variable indicates the difference between the effect of a cross-party interaction and that of an interaction (either cross-party or same-party).}

\begin{figure*}[!htbp]
\centering
\includegraphics[width=\textwidth]{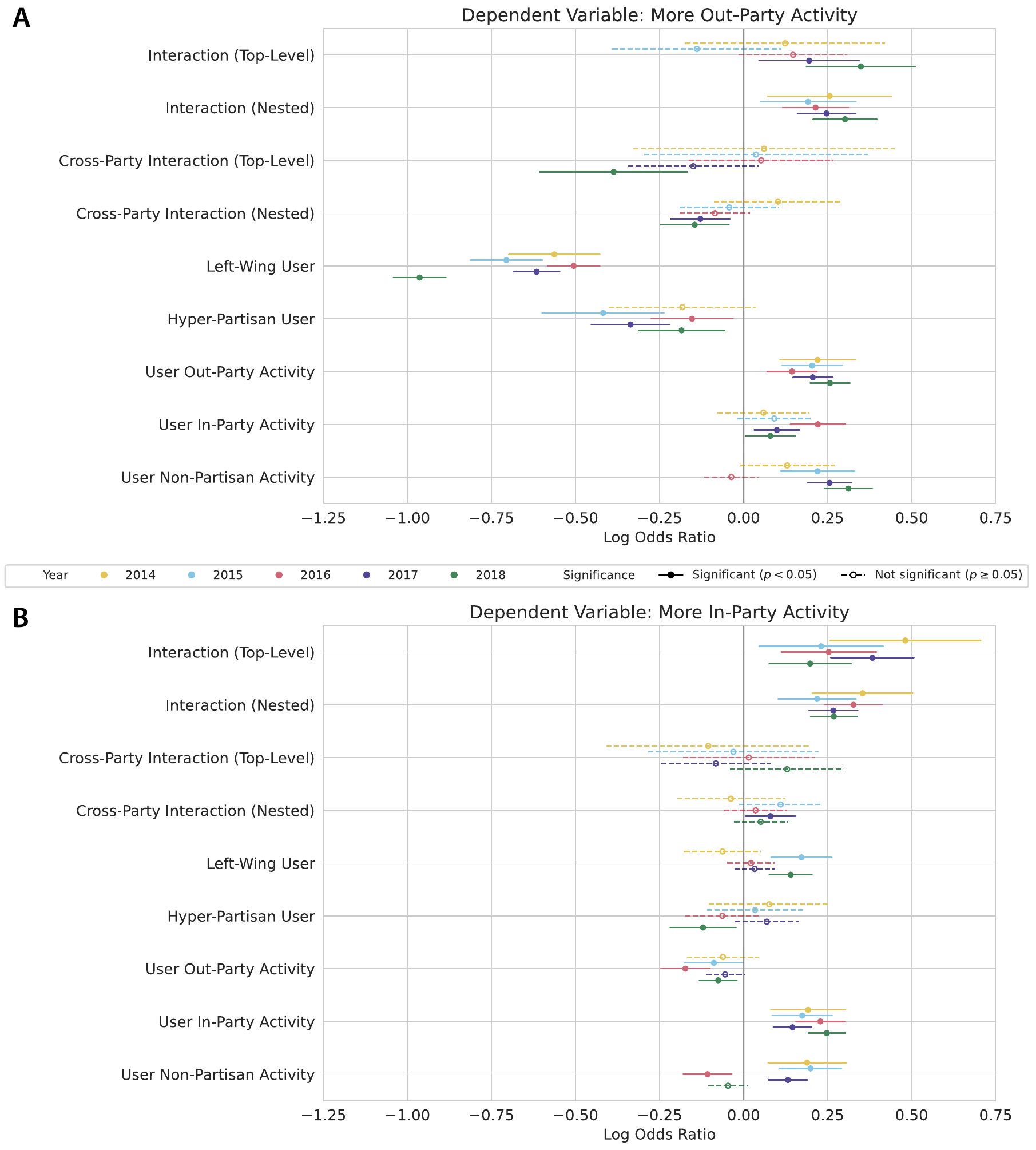}
\caption{Coefficients (i.e., log odds ratio) of the \newrev{difference-in-effects} regression model for predicting \rev{A) increased out-party subreddit activity, and B) increased in-party subreddit activity}, \newrev{where the coefficient of a \emph{Cross-Party Interaction} variable indicates the difference between the effect of a cross-party interaction and that of an interaction (either cross-party or same-party)}.
Each row corresponds to an independent variable. Each point on the row corresponds to the estimated coefficient of the independent variable in one year's model\rev{, and the bar underneath indicates the 95\% confidence interval of the estimate}. A solid circle \rev{with a solid line} indicates that the independent variable has a significant correlation with the dependent variable in the corresponding year, while a hollow circle \rev{with a dotted line} indicates a non-significant one. A coefficient above 0 indicates that the independent variable has a positive correlation with the dependent variable, while a coefficient below 0 indicates a negative one. 
}
\label{fig:res-diff}
\end{figure*}

\newrev{The results are shown in Figure~\ref{fig:res-diff}. Overall, we do not see a significant difference in effects that is consistent across all inspected years. While there does seem to be a growing trend of the difference over time where a cross-party interaction more and more negatively predicts increased out-party activity than an average interaction, it is unclear if we can interpret it as a growing backfire effect, as the growth in difference is mainly driven by an increasingly positive association between a same-party interaction and increased out-party activity, rather than an increasingly negative association between a cross-party interaction and increased out-party activity (Figure~\ref{fig:res}).}

\spara{Ratio models}
\newrev{We fit another regression model to inspect the association between an interaction and increased ratio of out-party activity in the after period, as opposed to increased absolute activity that we inspect in the main model. More specifically, we have \emph{More Out-Party Activity Ratio} as the dependent variable: let $a_o$ denote out-party subreddit activity and $a_i$ denote in-party subreddit activity, the dependent variable takes value 1 if the $a_o/(a_o+a_i)$ ratio is higher in the after period than in the before period, and 0 otherwise.}

\begin{figure*}[!htbp]
\centering
\includegraphics[width=\textwidth]{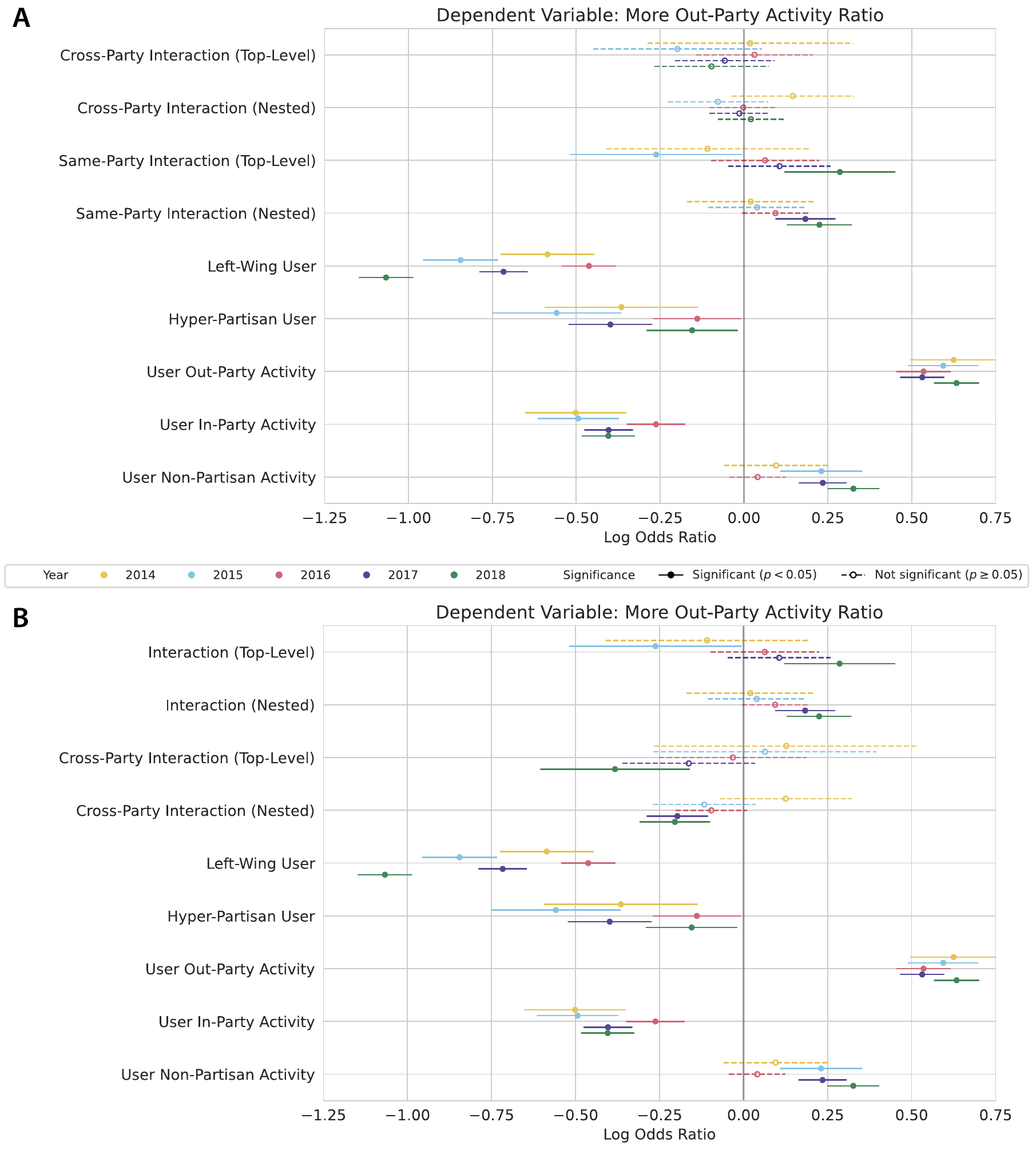}
\caption{\newrev{Coefficients (i.e., log odds ratio) of the A) absolute-effect regression model, and B) difference-in-effects regression model for predicting increased ratio of out-party subreddit activity.}
Each row corresponds to an independent variable. Each point on the row corresponds to the estimated coefficient of the independent variable in one year's model\rev{, and the bar underneath indicates the 95\% confidence interval of the estimate}. A solid circle \rev{with a solid line} indicates that the independent variable has a significant correlation with the dependent variable in the corresponding year, while a hollow circle \rev{with a dotted line} indicates a non-significant one. A coefficient above 0 indicates that the independent variable has a positive correlation with the dependent variable, while a coefficient below 0 indicates a negative one. 
}
\label{fig:ratio}
\end{figure*}

\newrev{The results of respectively the absolute-effect variant (where we have independent variables of \emph{Same-Party Interaction} and \emph{Cross-Party Interaction}) and the difference-in-effects variant (where we have independent variables of \emph{Interaction} and \emph{Cross-Party Interaction}) of this ratio model are shown in Figure~\ref{fig:ratio}. We mostly see similar patterns as those in our main results: following a cross-party interaction, there is no clear evidence of either a depolarization effect that increases out-party activity ratio, or a backfire effect that decreases out-party activity ratio (Figure~\ref{fig:ratio}A). Again, a cross-party interaction seems to be more and more negatively predicting increased out-party activity ratio over time than an average interaction (Figure~\ref{fig:ratio}B), but the growing difference comes from an increasingly positive association between a same-party interaction and increased out-party activity ratio.}

\newrev{Note that for fitting this ratio model, we chose to exclude users who had no subreddit activity in the after period to avoid zero-division issues (while they all had nonzero subreddit activity in the before period because they had a valid leaning). After the filtering, the remaining number of users is 5,209 out of 7,867 in 2014, 7,996 out of 11,613 in 2015, 14,203 out of 19,002 in 2016, 20,677 out of 27,042 in 2017, and 20,758 out of 27,118 in 2018. Our findings are thus specific to this subset of users, and are not guaranteed to hold upon inclusion of zero-activity users via alternative approaches.}

\section{Appendix B: Political Subreddits}
\label{sec:pol-sub}
Table~\ref{tab:left-sub}, \ref{tab:np-sub}, and \ref{tab:right-sub} list respectively the left-leaning subreddits, the non-partisan subreddits, and the right-leaning subreddits we identified, along with their corresponding leaning scores.

\begin{table}[p]
    \centering
    \small
    \begin{tabular}{lr}
    \toprule
    Subreddit & Leaning Score \\
    \midrule
    \texttt{r/democrats} & -2 \\
    \texttt{r/EnoughLibertarianSpam} & -2 \\
    \texttt{r/hillaryclinton} & -2 \\
    \texttt{r/progressive} & -2 \\
    \texttt{r/BlueMidterm2018} & -2 \\
    \texttt{r/Enough\_Sanders\_Spam} & -2 \\
    \texttt{r/GunsAreCool} & -2 \\
    \texttt{r/EnoughTrumpSpam} & -2 \\
    \texttt{r/Impeach\_Trump} & -2 \\
    \texttt{r/Trumpgret} & -2 \\
    \texttt{r/LateStageCapitalism} & -2 \\
    \texttt{r/FULLCOMMUNISM} & -2 \\
    \texttt{r/Fuckthealtright} & -2 \\
    \texttt{r/ShitLiberalsSay} & -2 \\
    \texttt{r/COMPLETEANARCHY} & -2 \\
    \texttt{r/ChapoTrapHouse} & -2 \\
    \texttt{r/Political\_Revolution} & -2 \\
    \texttt{r/Liberal} & -2 \\
    \texttt{r/asianamerican} & -2 \\
    \texttt{r/conspiratard} & -2 \\
    \texttt{r/liberalgunowners} & -2 \\
    \texttt{r/socialism} & -2 \\
    \texttt{r/The\_Mueller} & -2 \\
    \texttt{r/Kossacks\_for\_Sanders} & -2 \\
    \texttt{r/SandersForPresident} & -2 \\
    \texttt{r/BasicIncome} & -2 \\
    \texttt{r/TrumpCriticizesTrump} & -1 \\
    \texttt{r/neoliberal} & -1 \\
    \texttt{r/askhillarysupporters} & -1 \\
    \texttt{r/BannedFromThe\_Donald} & -1 \\
    \texttt{r/WayOfTheBern} & -1 \\
    \texttt{r/MarchAgainstTrump} & -1 \\
    \texttt{r/jillstein} & -1 \\
    \texttt{r/Anarchism} & -1 \\
    \texttt{r/The\_Dotard} & -1 \\
    \texttt{r/atheism} & -1 \\
    \texttt{r/PoliticalHumor} & -1 \\
    \texttt{r/inthenews} & -1 \\
    \texttt{r/politics} & -1 \\
    \bottomrule
    \end{tabular}
    \caption{List of left-leaning subreddits.}
    \label{tab:left-sub}
\end{table}

\begin{table}[!htb]
    \centering
    \small
    \begin{tabular}{lr}
    \toprule
    Subreddit & Leaning Score \\
    \midrule
    \texttt{r/Ask\_Politics} & 0 \\
    \texttt{r/AnythingGoesNews} & 0 \\
    \texttt{r/PoliticalDiscussion} & 0 \\
    \texttt{r/WhitePeopleTwitter} & 0 \\
    \texttt{r/Bad\_Cop\_No\_Donut} & 0 \\
    \texttt{r/moderatepolitics} & 0 \\
    \texttt{r/AskALiberal} & 0 \\
    \texttt{r/DNCleaks} & 0 \\
    \texttt{r/law} & 0 \\
    \texttt{r/Economics} & 0 \\
    \texttt{r/POLITIC} & 0 \\
    \texttt{r/WikiLeaks} & 0 \\
    \texttt{r/economy} & 0 \\
    \texttt{r/news} & 0 \\
    \texttt{r/PoliticalVideo} & 0 \\
    \texttt{r/NeutralPolitics} & 0 \\
    \texttt{r/JusticeServed} & 0 \\
    \texttt{r/GaryJohnson} & 0 \\
    \texttt{r/CapitalismVSocialism} & 0 \\
    \texttt{r/ShitRConservativeSays} & 0 \\
    \bottomrule
    \end{tabular}
    \caption{List of non-partisan subreddits.}
    \label{tab:np-sub}
\end{table}

\begin{table}[!htb]
    \centering
    \small
    \begin{tabular}{lr}
    \toprule
    Subreddit & Leaning Score \\
    \midrule
    \texttt{r/Conservative} & 2 \\
    \texttt{r/The\_Donald} & 2 \\
    \texttt{r/conservatives} & 2 \\
    \texttt{r/uncensorednews} & 2 \\
    \texttt{r/AskTrumpSupporters} & 2 \\
    \texttt{r/altright} & 2 \\
    \texttt{r/AskThe\_Donald} & 2 \\
    \texttt{r/ShitPoliticsSays} & 2 \\
    \texttt{r/CCW} & 2 \\
    \texttt{r/Catholicism} & 2 \\
    \texttt{r/Republican} & 2 \\
    \texttt{r/progun} & 2 \\
    \texttt{r/Firearms} & 2 \\
    \texttt{r/climateskeptics} & 2 \\
    \texttt{r/guns} & 2 \\
    \texttt{r/Anarcho\_Capitalism} & 2 \\
    \texttt{r/CringeAnarchy} & 2 \\
    \texttt{r/Libertarian} & 2 \\
    \texttt{r/TumblrInAction} & 2 \\
    \texttt{r/HillaryForPrison} & 2 \\
    \texttt{r/Christianity} & 2 \\
    \texttt{r/MensRights} & 1 \\
    \texttt{r/DebateAltRight} & 1 \\
    \texttt{r/GoldandBlack} & 1 \\
    \texttt{r/POTUSWatch} & 1 \\
    \texttt{r/gunpolitics} & 1 \\
    \texttt{r/ProtectAndServe} & 1 \\
    \texttt{r/conspiracy} & 1 \\
    \bottomrule
    \end{tabular}
    \caption{List of right-leaning subreddits.}
    \label{tab:right-sub}
\end{table}

\end{document}